\documentclass[A4paper,notitlepage,superscriptaddress,longbibliography,aps,twocolumn]{revtex4-1}
\usepackage{graphicx}
\graphicspath{{figures/}}
\usepackage{amsfonts}
\usepackage{amssymb}
\usepackage{mathrsfs}
\usepackage{soul}
\usepackage[usenames, dvipsnames]{color}
\usepackage[colorlinks=true,citecolor=blue,urlcolor  = purple]{hyperref}

\usepackage{tikz}

\usepackage[mathcal]{euscript}

\usepackage{amsmath}

\newcommand{\Tr}{\operatorname{Tr}}

\newcommand{\sh}{\operatorname{sh}}

\newcommand{\ket}[1]{\left| #1 \right\rangle}

\newcommand{\ketbra}[2]{| #1 \rangle \langle #2 |}

\newcommand{\abs}[1]{\left| #1\right|}
\newcommand{\mean}[1]{\langle #1 \rangle}

\newcommand{\kommentar}[1]{}
\usepackage{xcolor}

\usepackage{amsthm}

\newtheorem*{lemma*}{Lemma}
\newtheorem*{corollary*}{Corollary}
\theoremstyle{remark}

\usepackage{graphicx}

\usepackage{float}
\usepackage{color}
\definecolor{npurple}{rgb}{0.3,0,0.6}

\definecolor{mygray}{gray}{0.6}

\begin{document}

\setstcolor{red}

\title{Entanglement dynamics of two mesoscopic objects with gravitational interaction}
\date{\today}

\author{H. Chau Nguyen}
\email{chau.nguyen@uni-siegen.de}
\affiliation{Naturwissenschaftlich-Technische Fakult\"at, Universit\"at Siegen, Walter-Flex-Stra{\ss}e 3, 57068 Siegen, Germany}

\author{Fabian Bernards}
\affiliation{Naturwissenschaftlich-Technische Fakult\"at, Universit\"at Siegen, Walter-Flex-Stra{\ss}e 3, 57068 Siegen, Germany}


\begin{abstract}
We analyse the entanglement dynamics of the two particles interacting through
gravity in the recently proposed experiments aiming at testing quantum
signatures for gravity~[Phy. Rev. Lett 119, 240401 \& 240402 (2017)]. We consider
the open dynamics of the system under  decoherence due to the
environmental interaction. We show that as long as the coupling
between the particles is strong, the system does indeed develop entanglement,
confirming the qualitative analysis in the original proposals. We show that the
entanglement is also robust against  stochastic fluctuations in
setting up the system. The optimal interaction duration for the experiment is
computed. A condition
under which one can prove the entanglement in a device-independent manner is also derived. 

\end{abstract}

\maketitle

\section{Introduction}
The unification of quantum mechanics and general relativity has been perceived
as one of the most important open problems of modern physics. Although a substantial theoretical effort has been made, there is not yet an agreement on a single theory of quantum gravity~\cite{kiefer2012a}. One of the main difficulties of the field is the lack of
experimental support~\cite{kiefer2012a}. As a result, in recent years a number of
proposals for searching the signature of quantum gravity in various contexts
have been made~\cite{hossenfelder2018a}. Among these proposals, a simple one
making use of  recent advances in manipulating mesoscopic quantum mechanical systems was proposed by Bose \emph{et al.}~\cite{bose2017a}, and independently by Marletto and Vedral~\cite{marletto2017a}. The proposed experiment has been referred to as the `BMV experiment'~\cite{christodolou2018a}. 

\begin{figure}[t!]
\vspace{30pt}
\begin{minipage}{0.20\textwidth}
\includegraphics[width=1\textwidth]{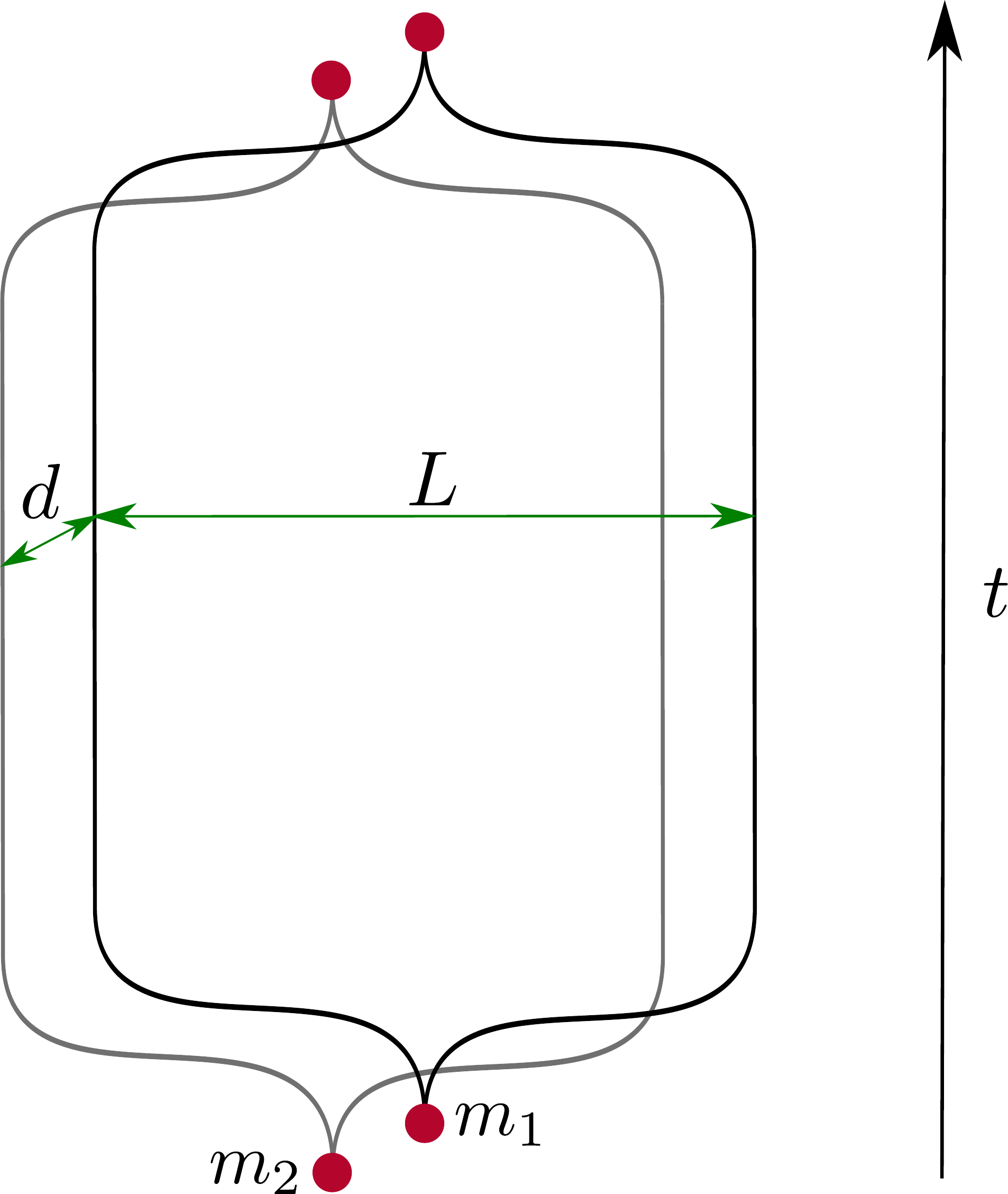} \\
(a)
\end{minipage}
\hspace{0.08\textwidth}
\begin{minipage}{0.16\textwidth}
\includegraphics[width=1\textwidth]{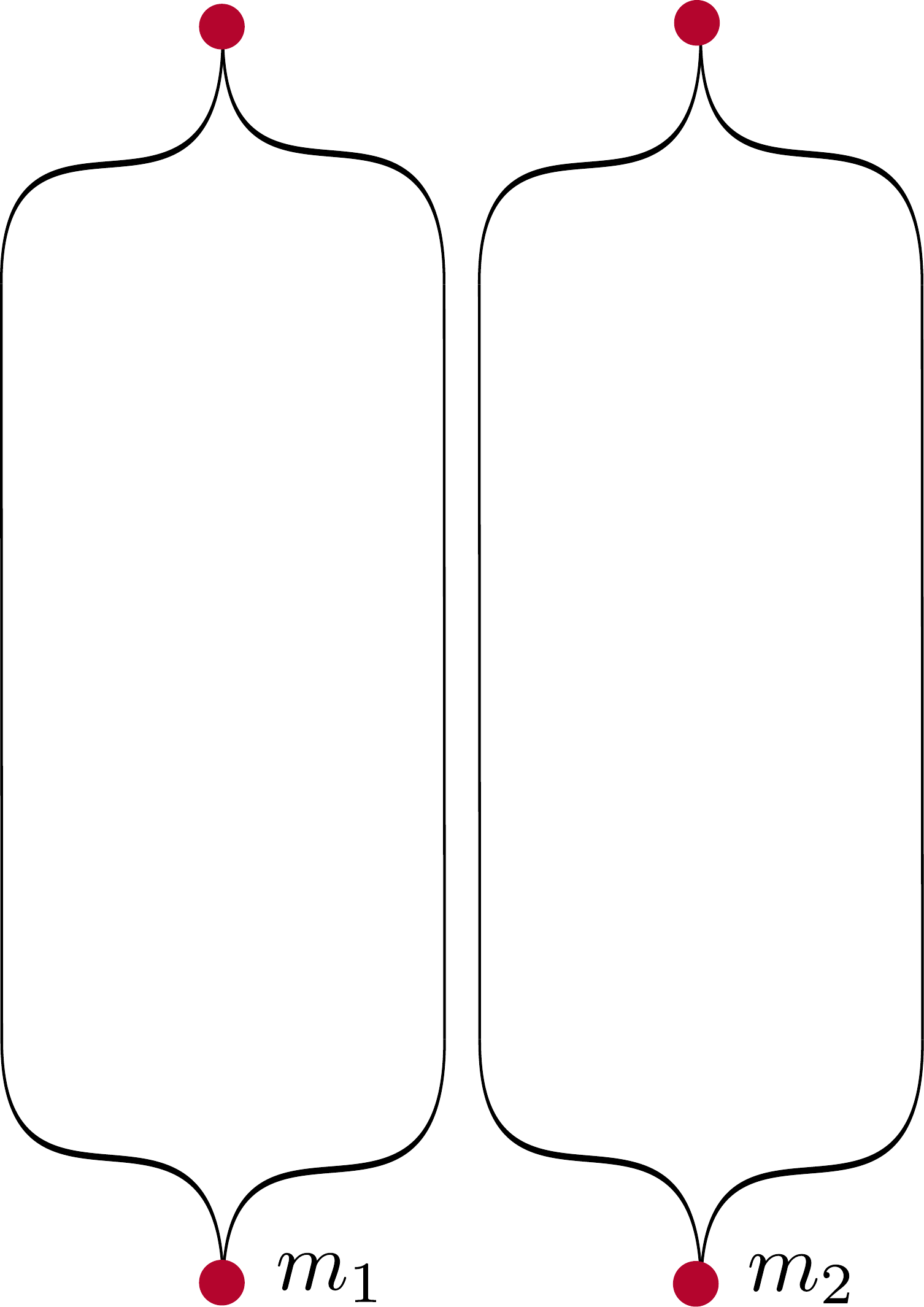}\\
(b)
\end{minipage}
\caption{The BMV experiment. (a) The symmetric setup: two particles are initially at distance $d$ from
each other. Each of the particle is then split into a superposition of two positions  at distance $L$ from each other in the direction orthogonal to their initial separation. Due to their
gravitational interaction the particles are expected to be entangled over time.
(b) The original setup: the particles are split into superpositions in parallel to their initial separation.}
\label{fig:bmv}
\end{figure}

We consider here a slightly different version of the BMV experiment, see
Figure~\ref{fig:bmv}(a). Two mesoscopic particles of masses $m_1$ and $m_2$  are
placed at distance $d$  from each other.  Each particle is then split into a superposition of two positions that are separated by a distance $L$ orthogonally to their initial separation. Based on recent advances in setting up mesoscopic systems in superposition~\cite{bose2018a,carley2018a,howl2017a,marshman2018a,qvarfort2018a,pino2018a}, the authors of Ref.~\cite{bose2017a} suggested as physically relevant quantities  $m_1 \approx m_2 \approx 10^{-8} \mathrm{kg}$, $d \approx 200 \mathrm{\mu m}$, and we can assume $L \gg d$.

In the original
proposal~\cite{bose2017a}, the particles are split into superpositions of positions in the same direction as their initial separation, see Figure~\ref{fig:bmv}(b). They thus have strong gravitational interaction only when the first particle is on the right, while the second particle is on the left. 
In this current setup the particles interact strongly whenever they are on the same side (left or right). Experimentally, it might be more challenging to
setup the system in this symmetric configuration; in particular, one may have to introduce a thin film between the
particles to keep the distance constant. We note that a thin film however may have an additional advantage.  It could help prevent taming interactions due to the van der Waals or the Casimir effects, which has been an obstacle for the original setup~\citep{bose2018a}. For convenience, we analyse this
symmetric setup, but the analysis can be easily adapted to the original
proposal.

Formally, the system can be modelled as a pair of spins, where states
$\ket{\uparrow}$ and $\ket{\downarrow}$ can be identified with the particles
being on the left and right, respectively. Due to the gravitational interaction
between the particles, if both particles are on the same sides ($\ket{\uparrow \uparrow}$ or $\ket{\downarrow \downarrow}$), the energy of the system is $-Gm_1m_2/d$, with $G$ being the gravitational constant. On the other hand, if they are on the opposite sides ($\ket{\uparrow \downarrow}$ or $\ket{\downarrow \uparrow}$), the energy is given by $-Gm_1m_2/\sqrt{L^2+d^2}$. Therefore, upto an irrelevant additive constant, the Hamiltonian can be modelled by
\begin{equation}
H= -\frac{\Delta}{2} \sigma_z \otimes \sigma_z,
\label{eq:hamiltonian}
\end{equation}
where $\sigma_z$ is one of the usual Pauli matrices and
\begin{equation}
\Delta= G m_1 m_2 \left( \frac{1}{d} - \frac{1}{\sqrt{L^2+d^2}} \right).
\end{equation}
Under the evolution induced by this Hamiltonian,
particles that are first given in the superpositions of being left and
right, $\ket{+}\ket{+}$, where
$\ket{+}=(\ket{\uparrow}+\ket{\downarrow})/\sqrt{2}$, should evolve into an
entangled state. Provided one can preserve the
coherence of the system long enough, such
entanglement is expected to be observable~\cite{bose2017a}.
In the actual physical setting, each particle carries an additional two-level
degree of freedom, which is then correlated with its positions (left or right)
during the spliting~\cite{bose2017a}. The result is that, after merging their
superpositions, the entanglement in the particle positions is eventually
transferred to the entanglement between these additional degrees of freedom and can be directly measured.

The entanglement between the
particles has been argued to be an evidence that the gravitational field is a quantum mechanical system~\cite{bose2017a,marletto2017a}. 
While this claim is still a subject of debate~\cite{hall2018a,anastopoulos2018a,reginatto2018a,christodolou2018a,christodolou2018b},
we do believe that the ability to entangle particles via their gravitational
interaction would greatly advance our understanding of the interface between quantum mechanics and gravity.  

Let $T$ be the decoherence time, the authors of Ref.~\cite{bose2017a} argued that the necessary condition to observe the entanglement is 
\begin{equation}
(\Delta T)/\hbar \sim \mathcal{O}(1).
\end{equation}
While this qualitative estimate is plausible, 
it is still important to analyse the noisy dynamics of the system in
detail to pinpoint the precise condition under which 
entanglement between the particles can be observed. Here we analyse the
details of the decoherence dynamics of the system.  
More importantly, we also consider
fluctuations of the experimental parameters. These
stochastic fluctuations in setting the parameters in the proposed
experiments imply that one has to average over the obtained
entangled states from run to run of the experiment, which results in a reduction of entanglement in the averaged state. While so far this has not been considered, it is also crucial to
the experiment, since entanglement can only be verified statistically through
multiple runs of the experiment. Our analysis shows that the
entanglement is rather robust. More precisely, we show that the entanglement indeed develops as long as $\Delta T / \hbar > 1$ if the fluctuations in setting the experimental parameters can be neglected. Moreover, we show that moderate stochastic errors in the experiment can also be tolerated. We discuss the optimal
interaction duration for the particles while they are in the superposition state and
find  a condition under which 
entanglement can be detected in a device-independent manner.
\section{The decoherence dynamics of the system}
The superposition of the positions of the particles is suppressed in the long time limit because of the environmental interaction. This is known as the decoherence process, which gives rise to our classical notion of position~\cite{schlosshauer2007a}. While the details of the decoherence process depend on the details of the environment, the system under consideration is sufficiently simple that it can be analysed with some minimal assumptions of the decoherence theory. Indeed, due to decoherence the system decays into a mixed state of positions (and not any other basis), so one can assume that the environment couples only to the position operator of the particles, which is $\sigma_z$  in this case~\cite{schlosshauer2007a}. The coupling Hamiltonian between one particle and the environment can be written generally as
\begin{equation}
H_{D} \propto \sigma_z \otimes R,
\label{eq:decoherence1}
\end{equation} 
where $R$ is an operator acting on the environment. This environmental coupling Hamiltonian~\eqref{eq:decoherence1} is such that if the initial reduced state of the particle is given by a $(2 \times 2)$ density matrix $a$, it will evolve in a way that its diagonal elements are constant, while its off-diagonal elements decay over time~\cite{schlosshauer2007a}. Assuming an exponential decay of the off-diagonal elements (known as coherence elements) for specificity, the state of the particle state at time $t$ is given by 
\begin{equation}
\rho_1(t)=
\begin{pmatrix}
a_{11} & a_{12} e^{-t} \\
a_{21} e^{-t} & a_{22}
\end{pmatrix},
\label{eq:decoherence1_form}
\end{equation} 
where we have used the dimensionless time $t$, defined by the physical time divided by the decoherence time $T$. While this exponential decaying of coherence is the case for the position decoherence due to the enviromental scattering by photons or air molecules~\cite{schlosshauer2007a}, other types of decoherence dynamics can also be considered with minimal adaptation. 

For the system of two particles without mutual interaction, we assume that their decoherence are independent from each other. If the system is first given in the state $a \otimes b$, where $a$ and $b$ are $(2 \times 2)$ density matrices of the first and the second particle, respectively, the density matrix of the whole system at time $t$ is then
\begin{equation}
\begin{pmatrix}
a_{11} & a_{12} e^{-t} \\
a_{21} e^{-t} & a_{22}
\end{pmatrix}
\otimes
\begin{pmatrix}
b_{11} & b_{12} e^{-t} \\
b_{21} e^{-t} & b_{22}
\end{pmatrix}.
\end{equation}
By linearity, the two-particle system first given in a $(4 \times 4)$ density matrix $c$ then evolves to
\begin{equation}
\begin{pmatrix}
c_{11} & c_{12}e^{- t} & c_{13}e^{- t} & c_{14} e^{-2 t} \\
c_{21} e^{-t} & c_{22} & c_{23} e^{-2 t} & c_{24} e^{- t} \\
c_{31} e^{- t} & c_{32} e^{-2 t} & c_{33} & c_{34} e^{- t} \\
c_{41} e^{- 2 t } & c_{42} e^{- t} & c_{43} e^{- t} & c_{44} 
\end{pmatrix}.
\label{eq:decoherence2}
\end{equation}


Let us now consider the interaction between the particles via the Hamiltonian~\eqref{eq:hamiltonian}. Importantly, the system Hamiltonian commutes with the environmental coupling~\eqref{eq:decoherence1}, rendering the total dynamics also exactly solvable regardless of the details of the enviroment operator $R$. Indeed, as we transform to the interaction picture by substituting $\rho= U(t) \rho_I(t) U^\dagger (t)$, where $U(t)=e^{-i H t}$, we find that the interacting density matrix $\rho_I(t)$ follows the dynamics of two independent particles interacting only with the environment given by equation~\eqref{eq:decoherence2}.
Assuming that at $t=0$ the system is in the state $\ketbra{+}{+}\otimes \ketbra{+}{+}$, we find the density matrix of the system at time $t$ to be
\begin{equation}
\rho = 
\frac{1}{4}
\begin{pmatrix}
1 & e^{i \omega t - t} & e^{i \omega t - t} & e^{-2 t} \\
e^{-i \omega t -t} & 1 & e^{-2 t} & e^{-i \omega - t} \\
e^{-i \omega t - t} & e^{-2 t} & 1 & e^{-i\omega t - t} \\
e^{- 2 t } & e^{i \omega t - t} & e^{i \omega t - t} & 1 
\end{pmatrix}.
\label{eq:noiseless_state}
\end{equation}
Recall that we are using the dimensionless time $t$, and $\omega=\Delta T/\hbar$ is referred to as the (dimensionless) coupling of the system.

\begin{figure}[hbt]
\begin{minipage}{0.24\textwidth}
\begin{center}
(a) \\
\includegraphics[width=\textwidth]{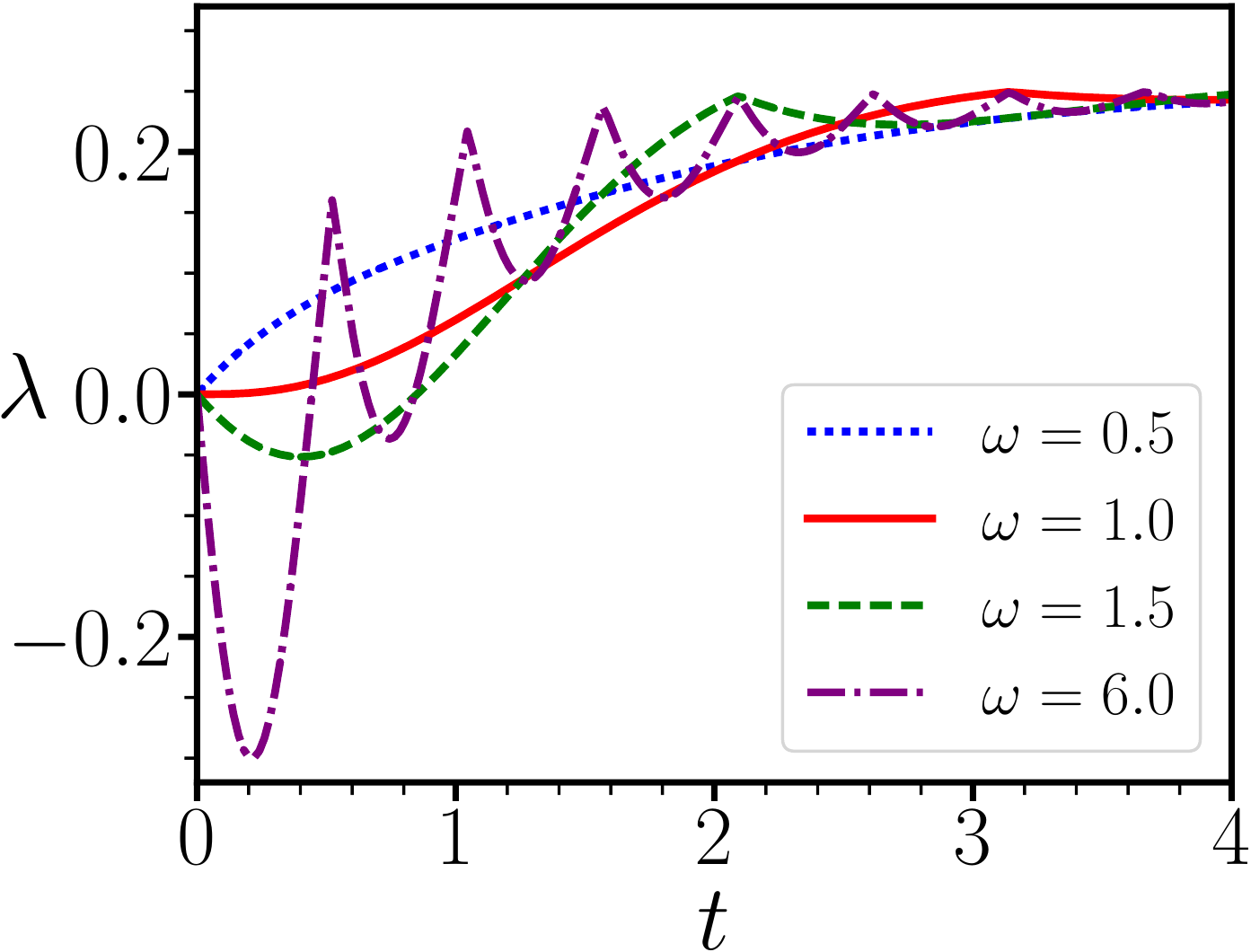}
\end{center}
\end{minipage}
\begin{minipage}{0.237\textwidth}
\begin{center}
(b) \\
\includegraphics[width=\textwidth]{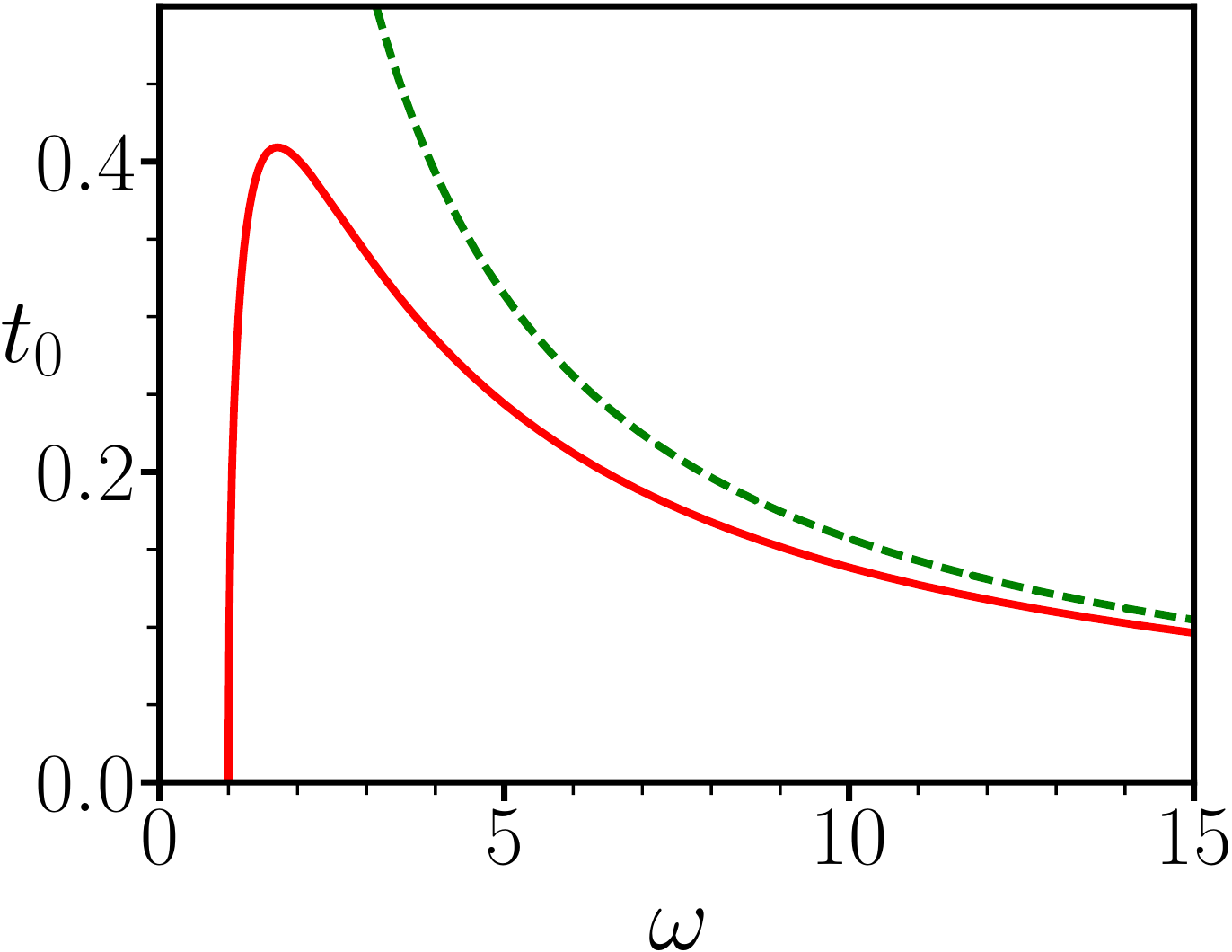}
\end{center}
\end{minipage}
\caption{(a) The smallest eigenvalue $\lambda$ of the partial transposition
of $\rho$ as a function of time.  (b) The optimal time where the two
particles are most entangled. The dashed-line indicates the asymptotic optimal
time in the limit of very strong couplings $\omega$ and no decoherence, given by $\pi/(2\omega)$.}
\label{fig:lambda}
\end{figure}

To analyse the entanglement dynamics in the density matrix~\eqref{eq:noiseless_state}, we use the positive partial transposition (PPT) criterion~\cite{peres1996a,horodecki1996a}. The smallest eigenvalue of the partial transposition of $\rho$ is found to be
\begin{equation}
\lambda= \frac{1}{2} e^{-t}(\sh t -  \abs{\sin \omega t}).
\label{eq:lambda}
\end{equation}
According to the PPT criterion, the two particles are entangled if and only if $\lambda < 0$. 
Figure~\ref{fig:lambda} (a) illustrates several different evolutions of $\lambda$ for different couplings $\omega$.
For $\omega<1$, $\lambda$ is positive and no entanglement develops. For
$\omega>1$, $\lambda$ becomes negative for certain times,
indicating that entanglement develops between the particles. This sharp
transition can be easily confirmed by analysing equation~\eqref{eq:lambda}. For
very large coupling parameters $\omega$, the particles can undergo entangled-disentangled oscillations. 
Obviously, the particles share the highest amount of entanglement during the
first phase of entanglement, where the effect of decoherence is still weak.
 To estimate the
optimal duration $t_0$ of the experiment, we find the first minimum of
$\lambda$. By considering the derivative of equation~\eqref{eq:lambda}, an
equation for $t_0$ can be found, namely
\begin{equation}
e^{-t_0}+\sin \omega t_0 -  \omega \cos \omega t_0 = 0,
\end{equation}
which is to be solved for the first positive time $t_0$. This
yields the  optimal duration for the experiment as a function of the coupling
parameter $\omega$. A plot of this function is presented in
Figure~\ref{fig:lambda} (b). For $1<\omega<1.8$, decoherence is strong
and the optimal time quickly increases with respect to the coupling strength
$\omega$ toward a maximum at $t_0 \approx 0.4$. For $\omega>1.8$, the internal evolution of the system dominates in the short time dynamics and the optimal time is similar to the time where the system achieves a Bell state when we ignore decoherence, $t_0 \approx \pi/(2\omega)$, which decreases as $\omega$ increases.   

\section{Stochastic fluctuations in preparing the experiment}
Let us consider now the fluctuations of the parameters of the experiment. If the
separation between the two positions of a particle $L$ is large in comparison to the typical wave
length of the electromagnetic environment, we can assume that the decoherence
time $T$ is not sensitive to this separation~\cite{schlosshauer2007a}. Moreover,
if $L$ is much larger than $d$, fluctuations in $L$ have only marginal effects on
$\Delta$. Thus, only fluctuations in two quantities are important: (a) fluctuations
in the minimal distance between the two particles $d$, which induce 
fluctuations in the coupling $\omega$ and (b) fluctuations in the interaction duration $t$.


To model the fluctuations, one can simply replace the deterministic values of
$\omega$ and $t$ by two gaussian random variables  $\omega + \xi_\omega
s_\omega$ and $t+ \xi_t s_t$, where $s_\omega$ and $s_t$ are
their standard deviations, and $\xi_\omega$ and $\xi_t$ are two standard
gaussian random variables. The state of the system averaged over all runs of the experiment would then be
\begin{equation}
\bar{\rho} =\mean{\rho(t)}_{\xi_t,\xi_\omega}.
\end{equation}

Assuming that the fluctuations are small, $t \gg s_t$, $\omega \gg
s_\omega$, one can expand their contributions in the phase and the damping terms
in $\rho$ to the first order in $s_\omega$ and $s_t$. Averaging over
the gaussian fluctuations in time $t$ and in coupling parameter $\omega$ then
yields the state
 \begin{align}
\bar{\rho} = \frac{1}{4}
\begin{pmatrix}
1  & a  & a  &b \\
\bar{a} &  1  &b  & \bar{a} \\
\bar{a} &  b&  1 & \bar{a}  \\
b&  a &  a&  1 
\end{pmatrix},
\label{eq:noisy_state}
\end{align}
with
$a = e^{i \omega t - t} e^{-\frac{1}{2} s_{\omega}^2 t^2 + \frac 12 s_t^2 (i \omega -
1)^2}$ and $b = e^{-2 t} e^{2 s_t^2}$.

\begin{figure}
\includegraphics[width=0.35\textwidth]{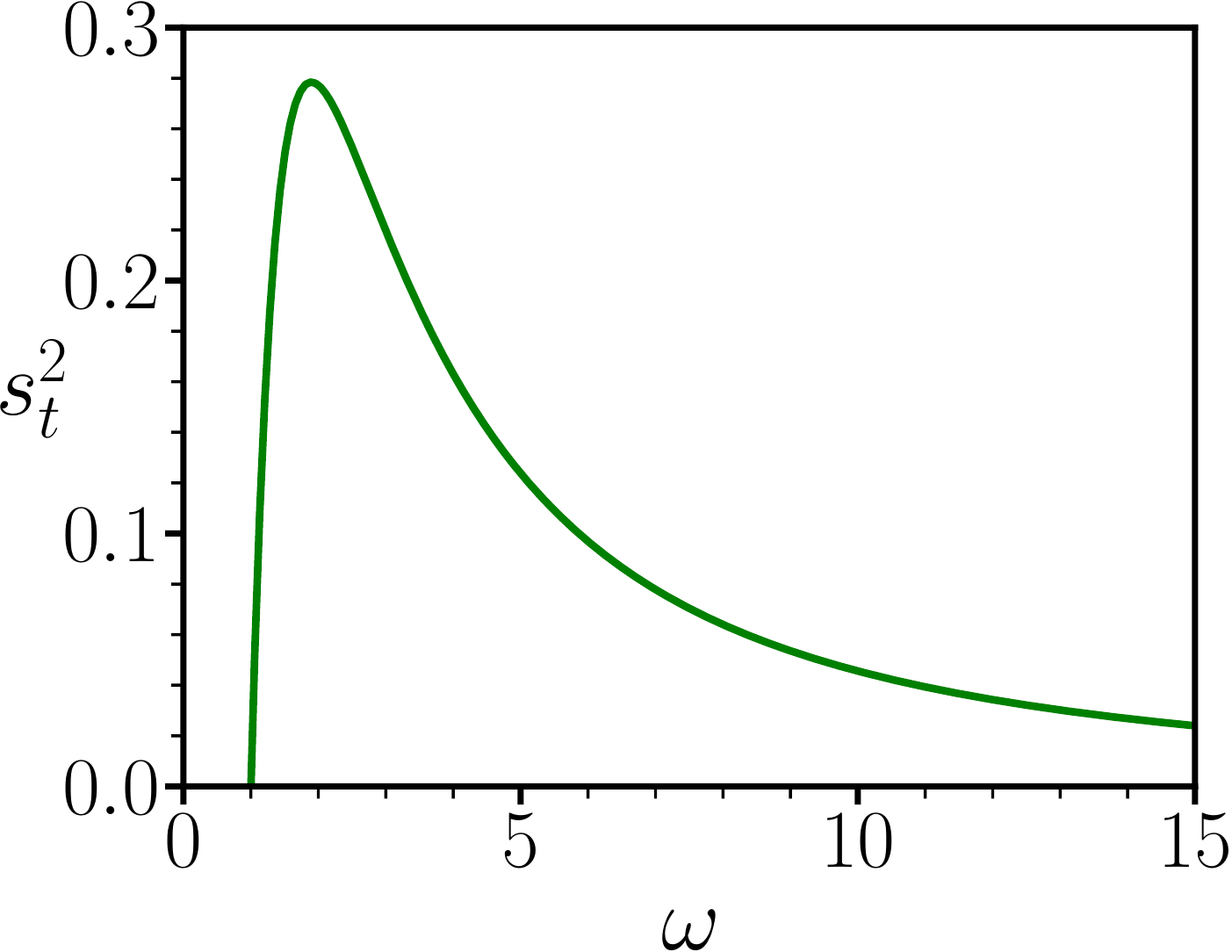}
\caption{Maximum fluctuation allowed in the interaction duration $s_t^2$ such that the entanglement can still be observed as a function of the coupling strength $\omega$.}
\label{fig:noisy_condition}
\end{figure}

To analyse the entanglement in the density operator, we again compute the smallest eigenvalue $\bar{\lambda}$ of the partial transposition of $\bar{\rho}$, which is
\begin{align}
\frac{1}{2} e^{-(t-s_t^2)}[ \sh (t-s_t^2)   - e^{- \frac{s_t^2}{2}(1+\omega^2)  - \frac{s_{\omega}^2}{2}t^2}
|\sin \omega (t-s_t^2)|].
\label{eq:lambda_bar}
\end{align}
By sending $s_\omega$ and $s_t$ to zero, we can easily recover
equation~\eqref{eq:lambda}. Note that
one should not extrapolate this formula to $t<s_t^2$, since in this regime
the fluctuations in time extrapolate the decoherence dynamics backward into
negative time, which is not physical. 
One then finds that the entanglement between the particles develops after $t > s_t^2$ if and only if
\begin{equation}
s_t^2 (1+s_t^2 s_\omega^2 + \omega^2) < 2 \ln \omega.
\end{equation}
To be consistent with the approximation, the higher order term $s_t^2 s_\omega^2$ should in fact be ignored. We thus obtain the condition
\begin{equation}
s_t^2 < \frac{2 \ln \omega}{1+\omega^2}.
\label{eq:sigma_t}
\end{equation} 
Remarkably, $s_\omega$ is absent in this condition. 
This means that if $\omega>1$, and if the
duration of interaction is well-controlled ($s_t \approx 0$), the entanglement persists despite
arbitrary fluctuations in $\omega$. On the other hand, equation~\eqref{eq:sigma_t} does pose a bound on the maximal standard deviation $s_t$ allowed, which is plotted in
Figure~\ref{fig:noisy_condition}. Interestingly, this indicates that if the
interaction strength is strong, one has to control the time more accurately;
on the other hand, in the intermediate regime, $s_t$ can vary to a large
extent.  If the interaction time can be precisely controlled by an atomic clock,
this accuracy can be easily achieved. In reality, the accuracy of the
interaction duraction in the actual experiment could be much less than the scale
of atomic clocks due to various difficulties in setting up the superposition
configuration for each particle. In particular, it is
known~\cite{baym2008a,mari2016a,belenchia2017a} that such a process should not
be too fast, otherwise gravitational radiations would interfere with the system
causing further decoherence effects. 
Yet, we expect that even in this case condition~\eqref{eq:sigma_t} can be easily achieved in reality.

\section{Violation of the CHSH inequality}
While the entanglement in the density operator can be demonstrated by state tomography or certain entanglement witnesses~\cite{guehne2009a}, it is generally desirable to demonstrate the entanglement in a device-independent way~\cite{brunner2014a}. 
This can be done by demonstrating a violation of the so-called Clauser-Horne-Shimony-Holt (CHSH) inequality~\cite{brunner2014a}. 
Suppose two parties, Alice and Bob, each of whom owns one particle of the pair.
Consider a situation where Alice performs either one of two measurements
$A_1,A_2$ on her particle while Bob performs either measurements $B_1$ or $B_2$
on his particle. Each measurement has only two outcomes $\pm 1$. If one
constrains that the system satifies the so-called assumption of local realism~\cite{einstein1935a,bell1964a}, then it is easy to show that
\begin{equation}
\abs{\mean{A_1B_1} + \mean{A_1B_2}+\mean{A_2B_1}-\mean{A_2B_2}} \le 2.
\label{eq:chsh}
\end{equation}
It has been repeatedly demonstrated in experiments that the CHSH inequality is
violated in quantum mechanics~\cite{brunner2014a}. This shows that quantum
mechanics is not compatible with the assumption of local realism, on
which the CHSH inequality~\eqref{eq:chsh} is based. What is relevant to us in
the current context is that in order to violate the CHSH
inequality~\eqref{eq:chsh} the state must be entangled. Notice that in order to
demonstrate the violation of the CHSH inequality, we only need the statistics of
the measurements  $A_1,A_2$ and $B_1,B_2$ in the experiment. The details how
such measurements are setup or how they are described mathematically are irrelevant~\cite{brunner2014a}.
In that sense, one can prove entanglement between particles in a device-independent way.

For simplicity  we ignore the fluctuations in the experimental parameters in
this section so that the density operator is of the simple
form as in equation~\eqref{eq:noiseless_state}. To see whether $\rho$ violates the CHSH
inequality for certain measurement settings, we make use of the criterion described in~Ref.~\cite{horodecki1995a}. To this end, we consider the correlation matrix $T_{ij}=\Tr [\rho \sigma_i \otimes \sigma_j]$, where $\sigma_i$ and $\sigma_j$ for $1 \le i,j \le 3$ are the Pauli matrices. It turns out that the correlation matrix $T$ for $\rho$ is degenerate with singular values $s_1= e^{-2t}$ and $s_2 =s_3= e^{-t} \abs{\sin \omega t}$. Then according to~\cite{horodecki1995a}, the state $\rho$ violates the CHSH inequality if and only if either $s_1^2 + s_2^2 > 1$ or $2 s_2^2 > 1$.
Solving these two inequalities numerically, we find that the system can
violate the
CHSH inequality if and only if
\begin{equation}
\omega > 4.19135.
\end{equation}
While this bound is significantly larger than the threshold of the coupling for
the systems to be entangled over time ($\omega > 1$), it is still in the same order of
magnitude. Thus once one can prove the entanglement of the particles, one is also
close to proving it in a device-independent manner.  
\section{Conclusion}

In this work we have analysed  the entanglement dynamics of the two particles in
the BMV experiment in detail. We showed that the system entangles as long as the coupling between the particles is strong, $\Delta T/\hbar
>1$, and the parameters are setup precisely.
Fluctuations in the parameters that arise from
setting up the experiment from run to run were then considered.  The entanglement turns out robust against the decoherence for 
some time and also against stochastic fluctuations. Moreover, we discuss the
optimal duration of the gravitational interaction while the particles are in a
superposition state. Also, we identify a condition under which one can detect
entanglement in a device-independent manner using the CHSH inequality.  We would like to mention that recently a similar detailed analysis of the entanglement dynamics has been made for another setup of the experiment~\cite{krisnanda2019a}. We hope
that together these analyses provide useful inputs for a realisation of the experiment in the near future.



\begin{acknowledgements} 
We thank Otfried G\"uhne
for helpful discussions and advice.
This work was supported by the DFG and the ERC (Consolidator Grant
683107/TempoQ) and the House of Young Talents Siegen. CN also acknowledges 
the support by the Vietnam National Foundation for Science and Technology 
Development (NAFOSTED) under grant number 103.02-2015.48.
\end{acknowledgements}

\bibliography{gravity}

\begin{thebibliography}{27}%
\makeatletter
\providecommand \@ifxundefined [1]{%
 \@ifx{#1\undefined}
}%
\providecommand \@ifnum [1]{%
 \ifnum #1\expandafter \@firstoftwo
 \else \expandafter \@secondoftwo
 \fi
}%
\providecommand \@ifx [1]{%
 \ifx #1\expandafter \@firstoftwo
 \else \expandafter \@secondoftwo
 \fi
}%
\providecommand \natexlab [1]{#1}%
\providecommand \enquote  [1]{``#1''}%
\providecommand \bibnamefont  [1]{#1}%
\providecommand \bibfnamefont [1]{#1}%
\providecommand \citenamefont [1]{#1}%
\providecommand \href@noop [0]{\@secondoftwo}%
\providecommand \href [0]{\begingroup \@sanitize@url \@href}%
\providecommand \@href[1]{\@@startlink{#1}\@@href}%
\providecommand \@@href[1]{\endgroup#1\@@endlink}%
\providecommand \@sanitize@url [0]{\catcode `\\12\catcode `\$12\catcode
  `\&12\catcode `\#12\catcode `\^12\catcode `\_12\catcode `\%12\relax}%
\providecommand \@@startlink[1]{}%
\providecommand \@@endlink[0]{}%
\providecommand \url  [0]{\begingroup\@sanitize@url \@url }%
\providecommand \@url [1]{\endgroup\@href {#1}{\urlprefix }}%
\providecommand \urlprefix  [0]{URL }%
\providecommand \Eprint [0]{\href }%
\providecommand \doibase [0]{http://dx.doi.org/}%
\providecommand \selectlanguage [0]{\@gobble}%
\providecommand \bibinfo  [0]{\@secondoftwo}%
\providecommand \bibfield  [0]{\@secondoftwo}%
\providecommand \translation [1]{[#1]}%
\providecommand \BibitemOpen [0]{}%
\providecommand \bibitemStop [0]{}%
\providecommand \bibitemNoStop [0]{.\EOS\space}%
\providecommand \EOS [0]{\spacefactor3000\relax}%
\providecommand \BibitemShut  [1]{\csname bibitem#1\endcsname}%
\let\auto@bib@innerbib\@empty
\bibitem [{\citenamefont {Kiefer}(2012)}]{kiefer2012a}%
  \BibitemOpen
  \bibfield  {author} {\bibinfo {author} {\bibfnamefont {C.}~\bibnamefont
  {Kiefer}},\ }\href@noop {} {\emph {\bibinfo {title} {Quantum gravity}}}\
  (\bibinfo  {publisher} {Cambridge University Press},\ \bibinfo {year}
  {2012})\BibitemShut {NoStop}%
\bibitem [{\citenamefont {Hossenfelder}(2018)}]{hossenfelder2018a}%
  \BibitemOpen
  \bibinfo {editor} {\bibfnamefont {S.}~\bibnamefont {Hossenfelder}},\ ed.,\
  \href@noop {} {\emph {\bibinfo {title} {Experimental search for quantum
  gravity}}}\ (\bibinfo  {publisher} {Springer Berlin},\ \bibinfo {year}
  {2018})\BibitemShut {NoStop}%
\bibitem [{\citenamefont {Bose}\ \emph {et~al.}(2017)\citenamefont {Bose},
  \citenamefont {Mazumdar}, \citenamefont {Morley}, \citenamefont {Ulbricht},
  \citenamefont {Toro\v{s}}, \citenamefont {Paternostro}, \citenamefont
  {Geraci}, \citenamefont {Barker}, \citenamefont {Kim},\ and\ \citenamefont
  {Milburn}}]{bose2017a}%
  \BibitemOpen
  \bibfield  {author} {\bibinfo {author} {\bibfnamefont {S.}~\bibnamefont
  {Bose}}, \bibinfo {author} {\bibfnamefont {A.}~\bibnamefont {Mazumdar}},
  \bibinfo {author} {\bibfnamefont {G.~W.}\ \bibnamefont {Morley}}, \bibinfo
  {author} {\bibfnamefont {H.}~\bibnamefont {Ulbricht}}, \bibinfo {author}
  {\bibfnamefont {M.}~\bibnamefont {Toro\v{s}}}, \bibinfo {author}
  {\bibfnamefont {M.}~\bibnamefont {Paternostro}}, \bibinfo {author}
  {\bibfnamefont {A.~A.}\ \bibnamefont {Geraci}}, \bibinfo {author}
  {\bibfnamefont {P.~F.}\ \bibnamefont {Barker}}, \bibinfo {author}
  {\bibfnamefont {M.~S.}\ \bibnamefont {Kim}}, \ and\ \bibinfo {author}
  {\bibfnamefont {G.}~\bibnamefont {Milburn}},\ }\bibfield  {title} {\enquote
  {\bibinfo {title} {Spin entanglement witness for quantum gravity},}\
  }\href@noop {} {\bibfield  {journal} {\bibinfo  {journal} {Phys. Rev. Lett.}\
  }\textbf {\bibinfo {volume} {119}},\ \bibinfo {pages} {240401} (\bibinfo
  {year} {2017})}\BibitemShut {NoStop}%
\bibitem [{\citenamefont {Marletto}\ and\ \citenamefont
  {Vedral}(2017)}]{marletto2017a}%
  \BibitemOpen
  \bibfield  {author} {\bibinfo {author} {\bibfnamefont {C.}~\bibnamefont
  {Marletto}}\ and\ \bibinfo {author} {\bibfnamefont {V.}~\bibnamefont
  {Vedral}},\ }\bibfield  {title} {\enquote {\bibinfo {title} {Gravitational
  induced entanglement between two massive particles is sufficient evidence of
  quantum effects in gravity},}\ }\href@noop {} {\bibfield  {journal} {\bibinfo
   {journal} {Phys. Rev. Lett.}\ }\textbf {\bibinfo {volume} {119}},\ \bibinfo
  {pages} {240402} (\bibinfo {year} {2017})}\BibitemShut {NoStop}%
\bibitem [{\citenamefont {Christodolou}\ and\ \citenamefont
  {Rovelli}(2018{\natexlab{a}})}]{christodolou2018a}%
  \BibitemOpen
  \bibfield  {author} {\bibinfo {author} {\bibfnamefont {N.}~\bibnamefont
  {Christodolou}}\ and\ \bibinfo {author} {\bibfnamefont {C.}~\bibnamefont
  {Rovelli}},\ }\bibfield  {title} {\enquote {\bibinfo {title} {On the
  possibility of laboratory evidence for quantum superposition of
  geometries},}\ }\href@noop {} {\bibfield  {journal} {\bibinfo  {journal}
  {arXiv:1808.05842v1}\ } (\bibinfo {year} {2018}{\natexlab{a}})}\BibitemShut
  {NoStop}%
\bibitem [{\citenamefont {Bose}\ and\ \citenamefont
  {Morley}(2018)}]{bose2018a}%
  \BibitemOpen
  \bibfield  {author} {\bibinfo {author} {\bibfnamefont {S.}~\bibnamefont
  {Bose}}\ and\ \bibinfo {author} {\bibfnamefont {G.~V.}\ \bibnamefont
  {Morley}},\ }\bibfield  {title} {\enquote {\bibinfo {title} {Matter and spin
  superposition in vacuum experiment {(MASSIVE)}},}\ }\href@noop {} {\bibfield
  {journal} {\bibinfo  {journal} {arXiv:1810.07045v1}\ } (\bibinfo {year}
  {2018})}\BibitemShut {NoStop}%
\bibitem [{\citenamefont {Carney}\ \emph {et~al.}(2018)\citenamefont {Carney},
  \citenamefont {Stamp},\ and\ \citenamefont {Taylor}}]{carley2018a}%
  \BibitemOpen
  \bibfield  {author} {\bibinfo {author} {\bibfnamefont {D.}~\bibnamefont
  {Carney}}, \bibinfo {author} {\bibfnamefont {P.~C.}\ \bibnamefont {Stamp}}, \
  and\ \bibinfo {author} {\bibfnamefont {J.~M.}\ \bibnamefont {Taylor}},\
  }\bibfield  {title} {\enquote {\bibinfo {title} {Tabletop experiments for
  quantum gravity: a user's maunal},}\ }\href@noop {} {\bibfield  {journal}
  {\bibinfo  {journal} {arXiv:1807.11494v2}\ } (\bibinfo {year}
  {2018})}\BibitemShut {NoStop}%
\bibitem [{\citenamefont {Howl}\ \emph {et~al.}(2017)\citenamefont {Howl},
  \citenamefont {Hackerm\"{u}ller}, \citenamefont {Bruschi},\ and\
  \citenamefont {Fuentes}}]{howl2017a}%
  \BibitemOpen
  \bibfield  {author} {\bibinfo {author} {\bibfnamefont {R.}~\bibnamefont
  {Howl}}, \bibinfo {author} {\bibfnamefont {L.}~\bibnamefont
  {Hackerm\"{u}ller}}, \bibinfo {author} {\bibfnamefont {D.~E.}\ \bibnamefont
  {Bruschi}}, \ and\ \bibinfo {author} {\bibfnamefont {I.}~\bibnamefont
  {Fuentes}},\ }\bibfield  {title} {\enquote {\bibinfo {title} {Gravity in the
  quantum lab},}\ }\href@noop {} {\bibfield  {journal} {\bibinfo  {journal}
  {Adv. Phys.}\ }\textbf {\bibinfo {volume} {3}},\ \bibinfo {pages} {138184}
  (\bibinfo {year} {2017})}\BibitemShut {NoStop}%
\bibitem [{\citenamefont {Marshman}\ \emph {et~al.}(2018)\citenamefont
  {Marshman}, \citenamefont {Mazumdar}, \citenamefont {Morley}, \citenamefont
  {Barker}, \citenamefont {Hoekstra},\ and\ \citenamefont
  {Bose}}]{marshman2018a}%
  \BibitemOpen
  \bibfield  {author} {\bibinfo {author} {\bibfnamefont {R.}~\bibnamefont
  {Marshman}}, \bibinfo {author} {\bibfnamefont {A.}~\bibnamefont {Mazumdar}},
  \bibinfo {author} {\bibfnamefont {G.~W.}\ \bibnamefont {Morley}}, \bibinfo
  {author} {\bibfnamefont {P.~F.}\ \bibnamefont {Barker}}, \bibinfo {author}
  {\bibfnamefont {S.}~\bibnamefont {Hoekstra}}, \ and\ \bibinfo {author}
  {\bibfnamefont {S.}~\bibnamefont {Bose}},\ }\bibfield  {title} {\enquote
  {\bibinfo {title} {Mesoscopic interference for metric and curvature {(MIMAC)}
  and gravitational waves},}\ }\href@noop {} {\bibfield  {journal} {\bibinfo
  {journal} {arXiv:1807.10830v1}\ } (\bibinfo {year} {2018})}\BibitemShut
  {NoStop}%
\bibitem [{\citenamefont {Qvarfort}\ \emph {et~al.}(2018)\citenamefont
  {Qvarfort}, \citenamefont {Bose},\ and\ \citenamefont
  {Serafini}}]{qvarfort2018a}%
  \BibitemOpen
  \bibfield  {author} {\bibinfo {author} {\bibfnamefont {S.}~\bibnamefont
  {Qvarfort}}, \bibinfo {author} {\bibfnamefont {S.}~\bibnamefont {Bose}}, \
  and\ \bibinfo {author} {\bibfnamefont {A.}~\bibnamefont {Serafini}},\
  }\bibfield  {title} {\enquote {\bibinfo {title} {Mesoscopic entanglement from
  central potential interactions},}\ }\href@noop {} {\bibfield  {journal}
  {\bibinfo  {journal} {arXiv:1812.09776v1}\ } (\bibinfo {year}
  {2018})}\BibitemShut {NoStop}%
\bibitem [{\citenamefont {Pino}\ \emph {et~al.}(2018)\citenamefont {Pino},
  \citenamefont {Prat-Camps}, \citenamefont {Sinha}, \citenamefont
  {Venkatesh},\ and\ \citenamefont {Romero-Isart}}]{pino2018a}%
  \BibitemOpen
  \bibfield  {author} {\bibinfo {author} {\bibfnamefont {H.}~\bibnamefont
  {Pino}}, \bibinfo {author} {\bibfnamefont {J.}~\bibnamefont {Prat-Camps}},
  \bibinfo {author} {\bibfnamefont {K.}~\bibnamefont {Sinha}}, \bibinfo
  {author} {\bibfnamefont {B.~P.}\ \bibnamefont {Venkatesh}}, \ and\ \bibinfo
  {author} {\bibfnamefont {O.}~\bibnamefont {Romero-Isart}},\ }\bibfield
  {title} {\enquote {\bibinfo {title} {On-chip quantum interference of a
  superconducting microsphere},}\ }\href@noop {} {\bibfield  {journal}
  {\bibinfo  {journal} {Quant. Sci. Tech.}\ }\textbf {\bibinfo {volume} {3}},\
  \bibinfo {pages} {25001} (\bibinfo {year} {2018})}\BibitemShut {NoStop}%
\bibitem [{\citenamefont {Hall}\ and\ \citenamefont
  {Reginatto}(2018)}]{hall2018a}%
  \BibitemOpen
  \bibfield  {author} {\bibinfo {author} {\bibfnamefont {M.~J.~W.}\
  \bibnamefont {Hall}}\ and\ \bibinfo {author} {\bibfnamefont {M.}~\bibnamefont
  {Reginatto}},\ }\bibfield  {title} {\enquote {\bibinfo {title} {On two recent
  proposals for witnessing nonclassical gravity},}\ }\href@noop {} {\bibfield
  {journal} {\bibinfo  {journal} {J. Phys. A: Math. Theor.}\ }\textbf {\bibinfo
  {volume} {51}},\ \bibinfo {pages} {085303} (\bibinfo {year}
  {2018})}\BibitemShut {NoStop}%
\bibitem [{\citenamefont {Anastopoulos}\ and\ \citenamefont
  {Hu}(2018)}]{anastopoulos2018a}%
  \BibitemOpen
  \bibfield  {author} {\bibinfo {author} {\bibfnamefont {C.}~\bibnamefont
  {Anastopoulos}}\ and\ \bibinfo {author} {\bibfnamefont {B.~L.}\ \bibnamefont
  {Hu}},\ }\bibfield  {title} {\enquote {\bibinfo {title} {Comment on ``a spin
  entanglement witness for quantum gravity'' and on ``gravitationally induced
  entanglement between two massive particles is sufficient evidence of quantum
  effects in gravity''},}\ }\href@noop {} {\bibfield  {journal} {\bibinfo
  {journal} {arXiv:1804.11315v1}\ } (\bibinfo {year} {2018})}\BibitemShut
  {NoStop}%
\bibitem [{\citenamefont {Reginatto}\ and\ \citenamefont
  {Hall}(2018)}]{reginatto2018a}%
  \BibitemOpen
  \bibfield  {author} {\bibinfo {author} {\bibfnamefont {M.}~\bibnamefont
  {Reginatto}}\ and\ \bibinfo {author} {\bibfnamefont {M.~J.~W.}\ \bibnamefont
  {Hall}},\ }\bibfield  {title} {\enquote {\bibinfo {title} {Entanglemetn of
  quantum fields via classical gravity},}\ }\href@noop {} {\bibfield  {journal}
  {\bibinfo  {journal} {arXiv:1809.04989v1}\ } (\bibinfo {year}
  {2018})}\BibitemShut {NoStop}%
\bibitem [{\citenamefont {Christodolou}\ and\ \citenamefont
  {Rovelli}(2018{\natexlab{b}})}]{christodolou2018b}%
  \BibitemOpen
  \bibfield  {author} {\bibinfo {author} {\bibfnamefont {N.}~\bibnamefont
  {Christodolou}}\ and\ \bibinfo {author} {\bibfnamefont {C.}~\bibnamefont
  {Rovelli}},\ }\bibfield  {title} {\enquote {\bibinfo {title} {On the
  possibility of experimental detection of the discreteness of time},}\
  }\href@noop {} {\bibfield  {journal} {\bibinfo  {journal}
  {arXiv:1812.01542v2}\ } (\bibinfo {year} {2018}{\natexlab{b}})}\BibitemShut
  {NoStop}%
\bibitem [{\citenamefont {Schlosshauer}(2007)}]{schlosshauer2007a}%
  \BibitemOpen
  \bibfield  {author} {\bibinfo {author} {\bibfnamefont {M.}~\bibnamefont
  {Schlosshauer}},\ }\href@noop {} {\emph {\bibinfo {title} {Decoherence and
  the quantum-to-classical transition}}}\ (\bibinfo  {publisher} {Springer
  Berlin},\ \bibinfo {year} {2007})\BibitemShut {NoStop}%
\bibitem [{\citenamefont {Peres}(1996)}]{peres1996a}%
  \BibitemOpen
  \bibfield  {author} {\bibinfo {author} {\bibfnamefont {A.}~\bibnamefont
  {Peres}},\ }\bibfield  {title} {\enquote {\bibinfo {title} {Separability
  criterion for density matrices},}\ }\href@noop {} {\bibfield  {journal}
  {\bibinfo  {journal} {Phys. Rev. Lett.}\ }\textbf {\bibinfo {volume} {77}},\
  \bibinfo {pages} {1413} (\bibinfo {year} {1996})}\BibitemShut {NoStop}%
\bibitem [{\citenamefont {Horodecki}\ \emph {et~al.}(1996)\citenamefont
  {Horodecki}, \citenamefont {Horodecki},\ and\ \citenamefont
  {Horodecki}}]{horodecki1996a}%
  \BibitemOpen
  \bibfield  {author} {\bibinfo {author} {\bibfnamefont {M.}~\bibnamefont
  {Horodecki}}, \bibinfo {author} {\bibfnamefont {P.}~\bibnamefont
  {Horodecki}}, \ and\ \bibinfo {author} {\bibfnamefont {R.}~\bibnamefont
  {Horodecki}},\ }\bibfield  {title} {\enquote {\bibinfo {title} {Separability
  of mixed states: Necessary and sufficient conditions},}\ }\href@noop {}
  {\bibfield  {journal} {\bibinfo  {journal} {Phys. Lett. A}\ }\textbf
  {\bibinfo {volume} {223}},\ \bibinfo {pages} {1} (\bibinfo {year}
  {1996})}\BibitemShut {NoStop}%
\bibitem [{\citenamefont {Baym}\ and\ \citenamefont {Ozawa}(2008)}]{baym2008a}%
  \BibitemOpen
  \bibfield  {author} {\bibinfo {author} {\bibfnamefont {G.}~\bibnamefont
  {Baym}}\ and\ \bibinfo {author} {\bibfnamefont {T.}~\bibnamefont {Ozawa}},\
  }\bibfield  {title} {\enquote {\bibinfo {title} {Two-slit diffraction with
  highly charged particles: {Niels Bohr's} consistency argument that the
  electromagnetic field must be quantized},}\ }\href@noop {} {\bibfield
  {journal} {\bibinfo  {journal} {Proc. Natl. Asoc. Sci.}\ }\textbf {\bibinfo
  {volume} {106}},\ \bibinfo {pages} {3035--3040} (\bibinfo {year}
  {2008})}\BibitemShut {NoStop}%
\bibitem [{\citenamefont {Mari}\ \emph {et~al.}(2016)\citenamefont {Mari},
  \citenamefont {De~Palma},\ and\ \citenamefont {Giovannetti}}]{mari2016a}%
  \BibitemOpen
  \bibfield  {author} {\bibinfo {author} {\bibfnamefont {A.}~\bibnamefont
  {Mari}}, \bibinfo {author} {\bibfnamefont {G.}~\bibnamefont {De~Palma}}, \
  and\ \bibinfo {author} {\bibfnamefont {V.}~\bibnamefont {Giovannetti}},\
  }\bibfield  {title} {\enquote {\bibinfo {title} {Experiments testing
  macroscopic quantum superpositions must be slow},}\ }\href@noop {} {\bibfield
   {journal} {\bibinfo  {journal} {Sci. Rep.}\ }\textbf {\bibinfo {volume}
  {6}},\ \bibinfo {pages} {22777} (\bibinfo {year} {2016})}\BibitemShut
  {NoStop}%
\bibitem [{\citenamefont {Belenchia}\ \emph {et~al.}(2018)\citenamefont
  {Belenchia}, \citenamefont {Wald}, \citenamefont {Giacomini}, \citenamefont
  {Castro-Ruiz}, \citenamefont {Brukner},\ and\ \citenamefont
  {Aspelmeyer}}]{belenchia2017a}%
  \BibitemOpen
  \bibfield  {author} {\bibinfo {author} {\bibfnamefont {A.}~\bibnamefont
  {Belenchia}}, \bibinfo {author} {\bibfnamefont {R.}~\bibnamefont {Wald}},
  \bibinfo {author} {\bibfnamefont {F.}~\bibnamefont {Giacomini}}, \bibinfo
  {author} {\bibfnamefont {E.}~\bibnamefont {Castro-Ruiz}}, \bibinfo {author}
  {\bibfnamefont {C.}~\bibnamefont {Brukner}}, \ and\ \bibinfo {author}
  {\bibfnamefont {M.}~\bibnamefont {Aspelmeyer}},\ }\bibfield  {title}
  {\enquote {\bibinfo {title} {Quantum superposition of massive objects and the
  quantization of gravity},}\ }\href@noop {} {\bibfield  {journal} {\bibinfo
  {journal} {arXiv:1807.07015v1}\ } (\bibinfo {year} {2018})}\BibitemShut
  {NoStop}%
\bibitem [{\citenamefont {G{\"u}hne}\ and\ \citenamefont
  {T{\'o}th}(2009)}]{guehne2009a}%
  \BibitemOpen
  \bibfield  {author} {\bibinfo {author} {\bibfnamefont {O.}~\bibnamefont
  {G{\"u}hne}}\ and\ \bibinfo {author} {\bibfnamefont {G.}~\bibnamefont
  {T{\'o}th}},\ }\bibfield  {title} {\enquote {\bibinfo {title} {{Entanglement
  detection}},}\ }\href@noop {} {\bibfield  {journal} {\bibinfo  {journal}
  {Phys. Rep.}\ }\textbf {\bibinfo {volume} {474}},\ \bibinfo {pages} {1--75}
  (\bibinfo {year} {2009})}\BibitemShut {NoStop}%
\bibitem [{\citenamefont {Brunner}\ \emph {et~al.}(2014)\citenamefont
  {Brunner}, \citenamefont {Cavalcanti}, \citenamefont {Pironio}, \citenamefont
  {Scarani},\ and\ \citenamefont {Wehner}}]{brunner2014a}%
  \BibitemOpen
  \bibfield  {author} {\bibinfo {author} {\bibfnamefont {N.}~\bibnamefont
  {Brunner}}, \bibinfo {author} {\bibfnamefont {D.}~\bibnamefont {Cavalcanti}},
  \bibinfo {author} {\bibfnamefont {S.}~\bibnamefont {Pironio}}, \bibinfo
  {author} {\bibfnamefont {V.}~\bibnamefont {Scarani}}, \ and\ \bibinfo
  {author} {\bibfnamefont {S.}~\bibnamefont {Wehner}},\ }\bibfield  {title}
  {\enquote {\bibinfo {title} {Bell nonlocality},}\ }\href@noop {} {\bibfield
  {journal} {\bibinfo  {journal} {Rev. Mod. Phys.}\ }\textbf {\bibinfo {volume}
  {86}},\ \bibinfo {pages} {419--478} (\bibinfo {year} {2014})}\BibitemShut
  {NoStop}%
\bibitem [{\citenamefont {Einstein}\ \emph {et~al.}(1935)\citenamefont
  {Einstein}, \citenamefont {Podolsky},\ and\ \citenamefont
  {Rosen}}]{einstein1935a}%
  \BibitemOpen
  \bibfield  {author} {\bibinfo {author} {\bibfnamefont {A.}~\bibnamefont
  {Einstein}}, \bibinfo {author} {\bibfnamefont {B.}~\bibnamefont {Podolsky}},
  \ and\ \bibinfo {author} {\bibfnamefont {N.}~\bibnamefont {Rosen}},\
  }\bibfield  {title} {\enquote {\bibinfo {title} {Can quantum-mechanical
  description of physical reality be considered complete?}}\ }\href@noop {}
  {\bibfield  {journal} {\bibinfo  {journal} {Phys. Rev.}\ }\textbf {\bibinfo
  {volume} {47}},\ \bibinfo {pages} {777--780} (\bibinfo {year}
  {1935})}\BibitemShut {NoStop}%
\bibitem [{\citenamefont {Bell}(1964)}]{bell1964a}%
  \BibitemOpen
  \bibfield  {author} {\bibinfo {author} {\bibfnamefont {J.}~\bibnamefont
  {Bell}},\ }\bibfield  {title} {\enquote {\bibinfo {title} {On the {Einstein
  Podolsky Rosen} paradox},}\ }\href@noop {} {\bibfield  {journal} {\bibinfo
  {journal} {Physics}\ }\textbf {\bibinfo {volume} {1}},\ \bibinfo {pages}
  {195--200} (\bibinfo {year} {1964})}\BibitemShut {NoStop}%
\bibitem [{\citenamefont {Horodecki}\ \emph {et~al.}(1995)\citenamefont
  {Horodecki}, \citenamefont {Horodecki},\ and\ \citenamefont
  {Horodecki}}]{horodecki1995a}%
  \BibitemOpen
  \bibfield  {author} {\bibinfo {author} {\bibfnamefont {R.}~\bibnamefont
  {Horodecki}}, \bibinfo {author} {\bibfnamefont {R.}~\bibnamefont
  {Horodecki}}, \ and\ \bibinfo {author} {\bibfnamefont {M.}~\bibnamefont
  {Horodecki}},\ }\bibfield  {title} {\enquote {\bibinfo {title} {Violating
  {Bell} inequality by mixed spin-$\frac{1}{2}$ states: necessary and
  sufficient condition},}\ }\href@noop {} {\bibfield  {journal} {\bibinfo
  {journal} {Phys. Lett. A}\ }\textbf {\bibinfo {volume} {200}},\ \bibinfo
  {pages} {340--344} (\bibinfo {year} {1995})}\BibitemShut {NoStop}%
\bibitem [{\citenamefont {Kristanda}\ \emph {et~al.}(2019)\citenamefont
  {Kristanda}, \citenamefont {Tham}, \citenamefont {Paternostro},\ and\
  \citenamefont {Paterek}}]{krisnanda2019a}%
  \BibitemOpen
  \bibfield  {author} {\bibinfo {author} {\bibfnamefont {T.}~\bibnamefont
  {Kristanda}}, \bibinfo {author} {\bibfnamefont {G.~Y.}\ \bibnamefont {Tham}},
  \bibinfo {author} {\bibfnamefont {M.}~\bibnamefont {Paternostro}}, \ and\
  \bibinfo {author} {\bibfnamefont {T.}~\bibnamefont {Paterek}},\ }\bibfield
  {title} {\enquote {\bibinfo {title} {Observable quantum entanglement due to
  gravity},}\ }\href@noop {} {\bibfield  {journal} {\bibinfo  {journal}
  {arXiv:1906.08808}\ } (\bibinfo {year} {2019})}\BibitemShut {NoStop}%
\end{thebibliography}%
\end{document}